%% file: fast.tex
\documentclass[Afour,sageh,times]{sagej_arxiv}

 \usepackage{graphicx}
 \input tex/header

\input tex/pf

\begin{document}

 \title{Highly Optimized Full-Core Reactor Simulations on Summit}

 \input tex/authors

 \input tex/abstract


\maketitle


 \input tex/introduction
 \input tex/algorithm

 \input tex/implementation

 \input tex/measurement   
 \input tex/performance

 \input tex/conclusion

 \input tex/acknowledge    

 \bibliographystyle{SageH}
 \bibliography{./bibs/emmd,./bibs/tim,./bibs/xobabko}



\end{document}

%% file: tex/header.tex
\usepackage{lineno,hyperref}
\modulolinenumbers[5]

\usepackage{tikz}
\usepackage{pgfplots}
\usepackage{pgfmath}
\usepackage[ruled,vlined,linesnumbered]{algorithm2e}
\DeclareTextCommandDefault{\textcopyright}{\textcircled{c}}
\DeclareTextCommandDefault{\textregistered}{\textcircled{%
      \check@mathfonts\fontsize\sf@size\z@\math@fontsfalse\selectfont R}}



%% file: tex/pf.tex
\newcommand{\dx}{\dftl{x}}

\newcommand{\dt}{\dftl{t}}

\newcommand{\be}{\begin{equation}}
\newcommand{\ee}{\end{equation}}
\newcommand{\bes}{\begin{equation*}}
\newcommand{\ees}{\end{equation*}}
\newcommand{\bse}{\begin{subequations}}
\newcommand{\ese}{\end{subequations}}

\newcommand{\uu}{\mathbf{u}}

\def\ee{{\hat {\underline e}}}

\def\dt{ \Delta t }
\def\dx{ \Delta x }

\def\scriptO{{{\it O}\kern -.42em {\it `}\kern + .20em}}
\def\RR{{{\rm l}\kern - .15em {\rm R} }}
\def\PP{{{\rm l}\kern - .15em {\rm P} }}
\def\L2{{{\sf L}^2}}
\def\H1{{{\sf H}^1}}

\def\PN2{{\PP_{N}-\PP_{N-2}}}

\def\complex{{{\rm C} \kern - .53em {\rm l} \kern + .38em}}

\def\a1{{ | \lambda_{\min} |}}

\def\l1{{   \lambda_{\min}  }}

\def\upt {\underline {\tilde p}}
\def\btu{{\tilde {\bf u}}}

\def\utu{{\tilde {\underline u}}}
\def\utb{{\tilde {\underline b}}}

\def\bu0{{\underline {\bf 0}}}

\def\bu{{\bf u}}

\def\bw{{\bf w}}
\def\bx{{\bf x}}

\def\At{{\tilde A}}

\def\Dh{{\hat D}}

\def\Oh{{\hat \Omega}}

\def\Jh{{\hat J}}

\def\ub{{\underline b}}

\def\up{{\underline p}}

\def\uu{{\underline u}}

\def\u0{{\underline 0}}
\def\n12{{n_{\frac{1}{2}}}}
\def\t12{{t_{\frac 1 2}}}
\def\n12{{n_{0.8}}}
\def\t12{{t_{0.8}}}

\def\udp{{\delta{\underline p}}}
\def\upb{{\bar {\underline p}}}

\newcommand{\pp}[2]{\frac{\partial #1}{\partial #2} }
\newcommand{\dd}[2]{\frac{d #1}{d #2} }


%% file: tex/authors.tex

\author{Paul Fischer,\affilnum{1,2,3}
        Elia Merzari,\affilnum{4,5}
        Misun Min,\affilnum{1}
        Stefan Kerkemeier,\affilnum{1}
        Yu-Hsiang Lan,\affilnum{1,2}
        Malachi Phillips,\affilnum{2}
        Thilina Rathnayake,\affilnum{2}
        April Novak,\affilnum{1}
        Derek Gaston,\affilnum{7}
        Noel Chalmers,\affilnum{8}
        Tim Warburton \affilnum{9}}

\corrauth{Misun Min,
Mathematics and Computer Science Division, Argonne National Laboratory, Lemont, IL 60439}
\email{mmin@mcs.anl.gov}

\affiliation{\affilnum{1} Mathematics and Computer Science Division, Argonne National Laboratory, Lemont, IL 60439\\
\affilnum{2} Department of Computer Science, University of Illinois at Urbana-Champaign, Urbana, IL 61801\\
\affilnum{3} Department of Mechanical Science and Engineering, University of Illinois at Urbana-Champaign, Urbana, IL 61801\\
\affilnum{4} Nuclear Science and Engineering Division, Argonne National Laboratory, Lemont, IL 60439\\
\affilnum{5} Department of Nuclear Engineering, Penn State, University Park, PA 16802\\ 
\affilnum{6} Department of Mechanical Engineering, Aristotle University, Thessaloniki, Greece\\
\affilnum{7} Idaho National Laboratory, Idaho Falls, ID 83415\\
\affilnum{8} AMD Research, Advanced Micro Devices Inc., Austin, TX 78735\\
\affilnum{9} Department of Mathematics, Virginia Tech, Blacksburg, VA 24061}

%% file: tex/abstract.tex
\begin{abstract}
Nek5000/RS is a highly-performant open-source spectral element code for 
simulation of incompressible and low-Mach fluid flow, heat transfer, and
combustion with a particular focus on turbulent flows in complex domains. It is
based on high-order discretizations that realize the same (or lower) cost per
gridpoint as traditional low-order methods.  State-of-the-art multilevel
preconditioners, efficient high-order time-splitting methods, and
runtime-adaptive communication strategies are built on a fast OCCA-based kernel
library, libParanumal, to provide scalability and portability across the
spectrum of current and future high-performance computing platforms. On Summit,
Nek5000/RS has recently achieved an milestone in the simulation of nuclear
reactors: the first full-core computational fluid dynamics simulations of
reactor cores, including pebble beds with $>$ 350,000 pebbles and 98M elements
advanced in less than 0.25 seconds per Navier-Stokes timestep.  
With carefully tuned algorithms, it is possible to simulate a single flow-through 
time for a full reactor core in less than six hours on all of Summit.

\end{abstract}

%% file: tex/introduction.tex
\section{Introduction}
We present first-ever large-eddy simulation of full reactor cores.
Our elliptic-operator kernels sustain $>$ 1 TFLOPS FP64 on V100s
and demonstarte  80\% parallel efficiency at $\approx 2.5$M gridpoints
per V100 with 0.1--0.25 second timesteps for Navier-Stokes problems
using 50--60B gridpoints on Summit.
We mesaure time-to-solution and provide scalability studies
on full-scale system of Summit.
We demonstrate {\it four-fold reduction} in solution time through
algorithmic advances based on our semi-implicit spectral element method
(SEM) approach with advanced preconditioning strategies utilizing mixed
precision.

%
%

 \begin{figure*}[t]
  \centering
  \includegraphics[width=1.\textwidth]{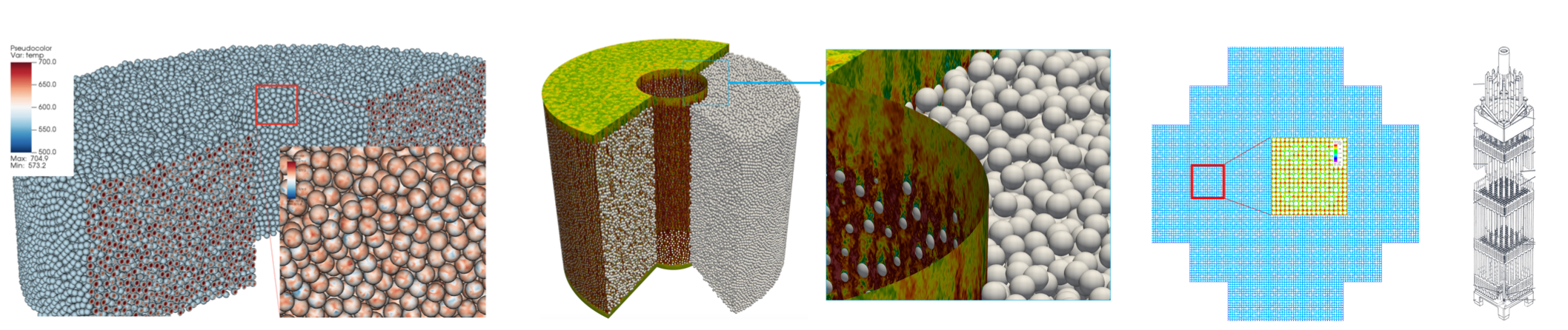}                  
 \caption{\label{fig:350k}
  Reactor simulations of turbulent flow performed on OLCF/Summit
  (left to right):
  MOOSE-NekRS simulation of start-up transient for 127,602 pebbles,
         temperatures in K;
  full core with 352,625 pebbles using an all-hex mesh comprising
  $E$=98,782,067 elements of order $N=8$;
  full core with 37 assemblies of 17$\times$17 pin bundles
  using 1 {\em billion} elements of order $N=5$;
  cut-away illustration of 17x17 bundle.}
\end{figure*}

We are interested in the modeling and simulation of nuclear reactor cores,
one the historic challenges associated with computing from its onset \cite{merzari2021, bajorek2016}.
The numerical simulation of nuclear reactors is an inherently multi-physics
problem involving fluid dynamics, heat transfer, radiation transport (neutrons)
and material and structural dynamics.

In particular we will focus here on the modeling of turbulent heat and mass
transfer in the core, which remains the most computationally expensive physics
to be solved at the continuum level \cite{grotzbach1999direct, merzari2017}. 
In fact, because of the complexity and
range of scales involved in reactor cores ($~1\mu m-10m$), simplifications
involving homogenization (e.g., treating the core effectively as porous medium)
or Reynolds averaging have been historically necessary \cite{novak2021, cleveland1986}. Such approaches, while
very powerful and useful for design, require modeling closures that are
expensive or impossible to develop and validate, especially at the appropriate
scale, and lead to design compromises in terms of operational margins and
economics \cite{roelofs2018thermal}.

In the past 10 years, simulation on leadership computing platforms has been
critical to improving informed analysis \cite{merzari2020wall}. In fact,
starting in 2009, turbulence-resolving simulations, such as direct numerical
simulation (DNS) or wall-resolved large eddy simulation (LES), of small
portions of the reactor core called fuel assemblies (e.g., $~100$ rods, $10 cm$
tall) have emerged.  These simulations have provided invaluable insight into
the physics involved in turbulent heat and mass transfer in reactor cores.
However, the simulation of a full reactor core has been elusive due to the
sheer size of the system (e.g., $~10,000$ rods, 2--3m tall), limiting the use
of such techniques, especially for multi-physics simulations that require
full core analysis.


Given the increased interest in advanced carbon-free nuclear fission reactors
worldwide, such turbulence-resolving full core simulations will be invaluable
to provide benchmark datasets for improved reduced-resolution models (including
advanced porous media models)\cite{}. This  will in turn lead to reduced uncertainty,
and ultimately to improved economics. In 2019 \cite{merzari2020toward} we
provided a timeline for full-core LES calculations and we estimated that a full
exascale machine would be necessary.  Thanks to several major innovations in
Nek5000/NekRS, this goal was recently achieved on pre-exascale machines.
We present here {\it the first full-core} computational fluid dynamics (CFD)
calculations to date.

In this article, we demonstrate that it is possible to realize a single
flow-through-time\footnote{The time required for a particle to be advected
by the flow through the entire domain.}
in just {\it six hours of wall-clock time} when running on all of
Summit (27648 NVIDIA V100s).   It is thus realistic to expect
that reactor designers could consider parametric analysis on
future exascale platforms, which will be substantially larger
than Summit.


To understand the transformative leap enabled by Nek5000/NekRS and in
particular NekRS, it is instructive to examine coupled simulations in pebble
beds. Before NekRS the largest  pebble bed calculations in the literature
involved of the order of $~1000$ pebbles with RANS \cite{van2018explicit} and
$~100$ pebbles with LES \cite{yildiz2020direct}. In this manuscript we describe
calculations that reach over the active section of the full core  for the
Mark-I Flouride Cooled High Temperature Reactor ($352,625$ pebbles)
\cite{andreades2016design}.

Cardinal \cite{merzari2021cardinal} couples NekRS to OpenMC, a Monte Carlo
solver for neutron transport, and to MOOSE, a multi-physics framework that has
been used to solve the complex physics involved in fuel performance.  The
coupled simulations of the Mark-I reactor have been performed on Summit with 6
MPI ranks per node corresponding to the 6 GPUs on each node. NekRS simulates
the flow and  temperature distribution around the pebbles on the GPUs; OpenMC
solves neutron transport on the CPU using 6 threads for each MPI rank; and
MOOSE solves unsteady conduction in the fuel pebbles on the CPU.  MOOSE can
also incorporate more complex fuel performance models (BISON) including fission
production and transport.  Coupled simulations on Summit have been performed for
pebble counts ranging from 11,145 pebbles on 100 nodes to 352,625 pebbles (full
core) on 3000 nodes.  In all cases, NekRS consumes $\approx$ 80\% of the
runtime. Figure~\ref{fig:350k}(left) represents  the temperature
distribution inside the pebbles and on the surface during a heat-up transient
(i.e., temperature starting from a constant temperature everywhere) for the
127,602 pebble case. These coupled simulations, which would have been
considered impossible only a few years ago, demonstrate that the majority
of the computational cost is in the thermal-fluids solve, which is the
focus of this study.



%% file: tex/algorithm.tex
\section{Algorithms}

Our focus is on the incompressible Navier-Stokes (NS) equations
for velocity ($\bu$) and pressure ($p$), 
\begin{eqnarray} \label{eq:nse}
\frac{D\bu}{Dt}\! :=\! \pp{\bu}{t} + \bu \cdot \nabla \bu \!=\!
\frac{1}{Re} \nabla^2 \bu - \nabla p , \;
\nabla \cdot \bu = 0, \;\;
\\[-4ex] \nonumber
\end{eqnarray} \normalsize
where $Re=UL/\nu \gg 1$ is the Reynolds number based on flow speed $U$,
length scale $L$, and viscosity $\nu$.
(The equation for temperature is similar to (\ref{eq:nse}) without
the pressure, which makes it much simpler.) From a computational
standpoint, the long-range coupling of the incompressibility constraint,
$\nabla \cdot \bu = 0$,
makes the pressure substep intrinsically communication intensive and
a major focus of our effort as it consumes 60-80\% of the run time.

NekRS is based on the spectral element method (SEM) \cite{pat84}, in which
functions are represented as $N$th-order polynomials on each of $E$ elements,
for a total mesh resolution of $n=EN^3$.  
While early GPU efforts for Nek5000 were on OpenACC ports~\cite{gong2015}
and \cite{min2015a}(for NekCEM), NekRS originates from two code suites, Nek5000
\cite{nek5000}  
and libParanumal \cite{warburton2019,warburton2019b}, 
which is a fast GPU-oriented library for high-order methods written in 
the open concurrent compute abstraction (OCCA) for cross-platform 
portability \cite{occa}.  

The SEM offers many advantages for this class of problems.  It accommodates
body-fitted coordinates through isoparametric mappings domain from the
reference element, $\Oh:=[-1,1]^3$ to the individual (curvilinear-brick)
elements $\Omega^e$.  On $\Oh$, solutions are represented in terms of
$N$th-order tensor-product polynomials,
\begin{eqnarray} \label{eq:field1} 
\left. \bu(\bx) \right|^{}_{\Omega^e} = \sum_{i=0}^N
\sum_{j=0}^N \sum_{k=0}^N \bu_{ijk}^{e}\,h_i(r)\,h_j(s)\,h_k(t), 
\end{eqnarray}
where the $h_i$s are stable nodal interpolants based on the 
Gauss-Lobatto-Legendre (GLL) quadrature points $(\xi_i,\xi_j,\xi_j)\in \Oh$
and $\bx=\bx^e(r,s,t)$ on $\Omega^e$.
This form allows all operator evaluations to be expressed as {\em fast tensor
contractions}, which can be implemented as BLAS3 operations\footnote{For
example, with $\Dh_{il} := \dd{h_l}{r} |^{}_{\xi_i}$,
the first derivative takes the form $u_{r,ijk}=\sum_l \Dh_{il} u_{ljk}$,
which is readily implemented as an $N_p \times N_p$ matrix times
an $N_p \times N_p^2$ matrix, with $N_p=N+1$.}
in only $O(N^4)$ work and $O(N^3)$ memory references \cite{dfm02,sao80}.
This low complexity is in sharp
contrast to the $O(N^6)$ work and storage complexity of the
traditional $p$-type FEM.  Moreover, hexahdral (hex) element function
evaluation is about six times faster per degree-of-freedom (dof) than
tensor-based tetrahedal (tet) operator evaluation \cite{moxey20}.
By diagonalizing one direction at a time, the SEM structure admits
fast block solvers for local Poisson problems in undeformed and (approximately)
in deformed elements, which serve as local smoothers for $p$-multigrid (pMG)
\cite{lottes05}.
$C^0$ continuity implies that the SEM is {\em communication minimal:}
data exchanges have unit-depth stencils, independent of $N$.
Finally, local $i$-$j$-$k$ indexing avoids much of the indirect addressing
associated with fully unstructured approaches, such that high-order SEM
implementations can
realize significantly higher throughput (millions of dofs per second, MDOFS)
than their low-order counterparts \cite{ceed_bp_paper_2020}.

The $O(N)$ computational intensity of the SEM brings direct benefits as its
rapid convergence (exponential in $N$) allows one to accurately simulate flows
with fewer gridpoints than lower-order discretizations.  Turbulence DNS
and LES require long-time simulations to reach statistically
steady states and to gather statistics.  For campaigns that can last weeks,
turbulence simulations require not only performant implementations but also
{\em efficient discretizations} that deliver high accuracy at low cost per
gridpoint.   Kreiss and Oliger \cite{kreiss72} noted that high-order methods
are important when fast computers enable long integration times.  To offset
cumulative dispersion errors, $e(t)\sim C t$, one must have $C \ll 1$ when $t
\gg 1$.  It is precisely in this regime that the asymptotic error behavior of
the SEM, $C$=$O(h^N)$, is manifest; it is more efficient to increase $N$
than to decrease the grid spacing, $h = O(E^{-\frac{1}{3}})$ (e.g., as in
Fig.  \ref{fig:art}(d)).

High-order incompressible NS codes comparable in scalability to 
NekRS include 
Nektar \cite{moxey20}, which employs the Nek5000 communication
           kernel, {\em gslib} and is based primarily on tets,  
libParanumal\cite{ChalmersKarakusAustinSwirydowiczWarburton2020},
NUMO\cite{numo},
deal.ii \cite{dealii}, and 
MFEM \cite{mfem}.  
The CPU-based kernel
performance for the latter two is comparable to Nek5000
\cite{ceed_bp_paper_2020}.  
    MFEM, which is a general purpose finite element library rather than a flow
solver, has comparable GPU performance because of the common collaboration
between the MFEM and Nek5000/RS teams that is part of DOE's Center for
Efficient Exascale Discretizations (CEED).  
    NUMO, which targets ocean modeling, is also OCCA based.
    libParanumal, which is also
under CEED, is at the cutting edge of GPU-based node performance on 
NVIDIA and AMD platforms \cite{ceed_special_issue2}.

Regarding large-scale NS solutions on Summit, the closest point of 
comparison is the spectral (Fourier-box) code of Ravikumar {\em et al.}
\cite{pkyeung19}, which achieves 14.24 s/step for $n=18432^3$ 
and $P=18432$ V100s (3072 nodes)---a throughput rate 
$R$=23.6 MDOFS (millions of points-per-[second-GPU]).
NekRS requires 0.24  s/step with $n=51$ billion and $P$=27648, for 
a throughput of $R$=7.62 MDOFS, for the full-reactor target problem 
of  Fig. \ref{fig:350k},
which means that NekRS is {\em within a factor of 3} of a dedicated
spectral code.  Traditionally, periodic-box codes, which do not
store geometry, support general derivatives, or use iterative solvers, 
run 10-20$\times$ faster than general purpose codes.
At this scale, however, the requisite all-to-alls for multidimensional 
FFTs of the spectral code place bisection-bandwidth burdens on the network 
that are not encountered with domain-decomposition approaches such as the SEM.
We further note that $> 1$ s/step is generally not practicable for
production turbulence runs, which typically require $10^5$--$10^6$
timesteps.

%% file: tex/implementation.tex

\begin{figure*}
{\setlength{\unitlength}{1.0in}
   \begin{picture}(6.500,1.60)(-.00,-.00)
      \put(0.07,0.00){\includegraphics[width=2.9in]{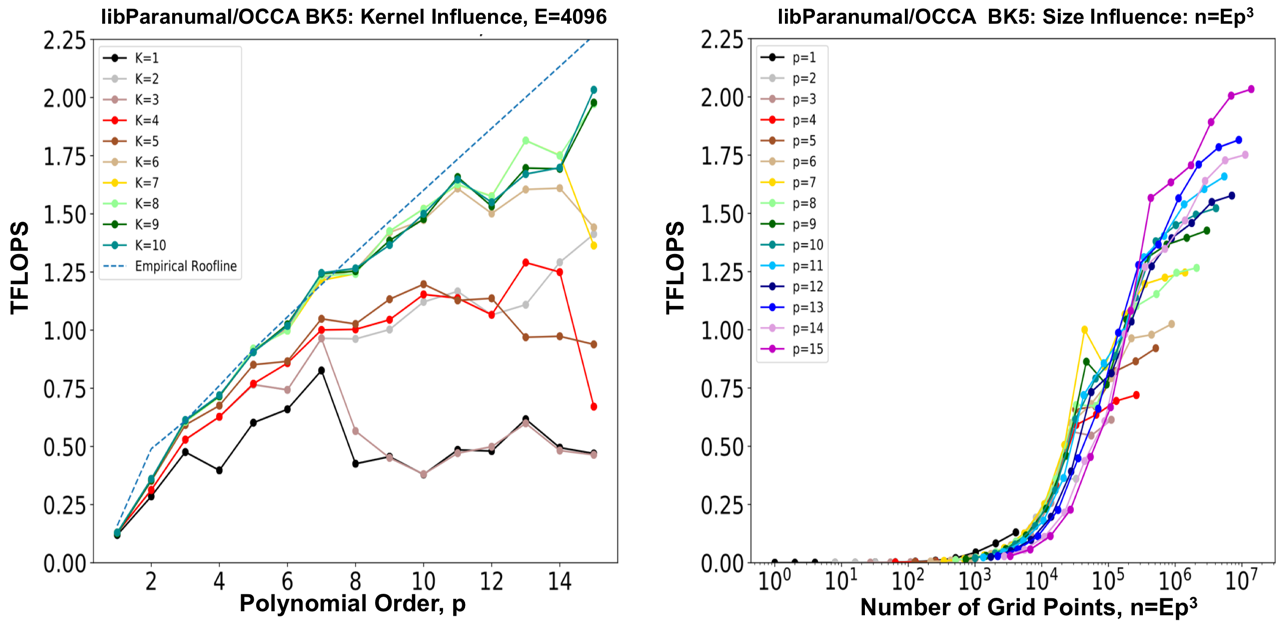}}
      \put(3.17,0.00){\includegraphics[width=3.1in]{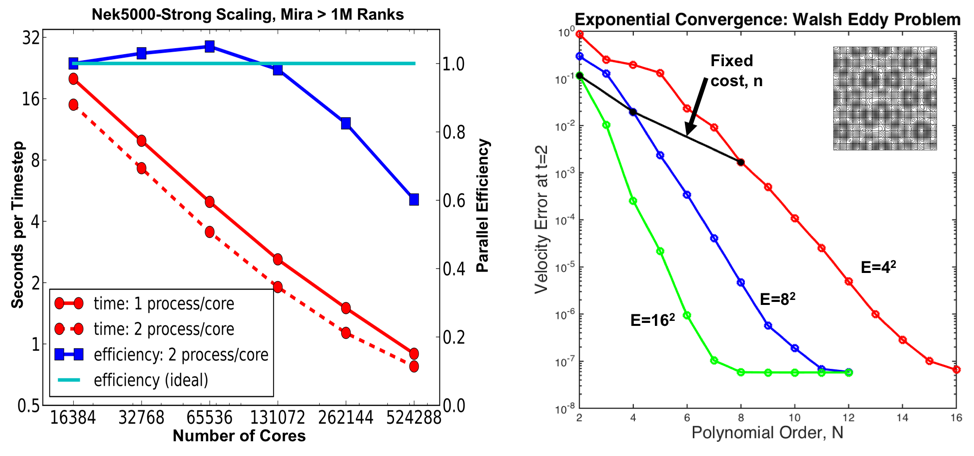}}
      \put(-.00,1.23){\tiny \bf (a)}
      \put(1.51,1.23){\tiny \bf (b)}
      \put(3.09,1.23){\tiny \bf (c)}
      \put(4.69,1.23){\tiny \bf (d)}
   \end{picture}}
\caption{\label{fig:art} \small
Basis for NekRS:
(a)--(b) Highly tuned OCCA-based Poisson kernel in libParanumal 
         saturating the V100 roof-line, 
(b) Poisson operator kernel performance (no communication)
    as a function of $p$ ($\equiv N$) and $n=Ep^3$,
(c) Nek5000 strong scaling to $>$ 1 million ranks, $n_{0.8} \approx 4000$
    on Mira, 
(d) illustration of high-order benefits for the advection-dominated Walsh
    problem studied in \cite{fischer17}. 
    At engineering accuracy, the 8th-order expansion is significantly 
    more accurate than the 2nd-order expansion for the same {\em cost}, $n$.  }
\end{figure*}
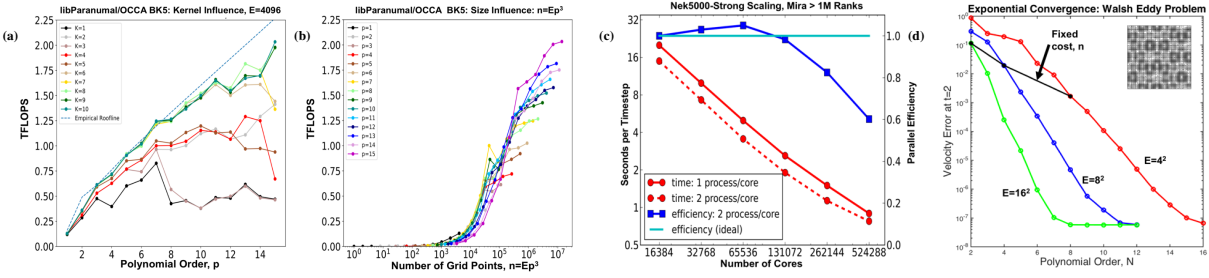

\def\tf{t_{\mbox{\tiny final}}}

\section{Implementation}

Our decision to use OCCA was motivated by the need to port {\em fast
implementations} to multiple-node architectures featuring Cuda, HIP, OpenMP,
etc., without major revisions to code structure.  Extensive tuning is 
required, which has been addressed in libParanumal (e.g.,
\cite{warburton2019,ceed_bp_paper_2020}).  Figures \ref{fig:art}(a) and (b)
show the performance of libParanumal on a single NVIDIA V100 for the local
spectral element Poisson operator (without communication) as a function of
kernel tuning (described in \cite{ceed_bp_paper_2020}), polynomial order $p$
($\equiv N$), and problem size $n$.   We see that the performance
saturates at the roofline for all cases when $E=4096$ in (a) and at 
$n \approx$ 250,000 for $p=$7--8 in (b).  As we discuss in the sequel,
because of communication overhead, we need $n/P \approx $ 2--3 million to
sustain 80\% parallel efficiency in the full Navier--Stokes case on Summit, 
where $P$ is the number of V100s employed in the simulation.

NekRS is based on libParanumal and leverages several of its
high-performance kernels, including features such as overlapped communication
and computation.   It also inherits decades of development work from Nek5000,
which has scaled to millions of ranks on Sequoia and Mira.  As shown in Fig.
\ref{fig:art}(c), the 80\% parallel efficiency mark for Nek5000 on Mira is at
$n/P\approx 4000$ (where $P$ is the number of ranks), in keeping with the strong-scale
analysis of \cite{fischer15} and the results obtained through cross-code comparisons
in \cite{ceed_bp_paper_2020} with the high-order deal.ii and MFEM libraries
\cite{dealii,mfem}.

Time advancement in NekRS is based on $k$th-order operator splitting
($k$=2 or 3) that decouples
(\ref{eq:nse}) into separate advection, pressure, and viscous substeps.
The nonlinear advection terms are advanced by using backward differencing 
(BDF$k$) applied to the material derivative (e.g., 
$D\bu/Dt \approx \frac{1}{2\dt}(3\bu^m-4\btu^{m-1}+\btu^{m-2})$ for $k$=2), 
where values $\btu^{m-j}$ along the characteristic are found by solving an 
explicit hyperbolic subproblem, $\pp{\bw}{t}= -\bu\cdot \nabla \bw$, 
on $[t^{m-1},t^m]$ \cite{patel18,maparo90}.  
For stability (only), the variable-coefficient hyperbolic problem is dealiased 
by using Gauss--Legendre quadrature on $N_q > N$ quadrature points \cite{johan13}.
The characteristics approach allows for Courant numbers,
CFL$:= \max_{\bx \in \Omega} \dt |u_i/\dx_i|$, significantly larger than unity
(CFL=2--4 is typical), thus reducing the required number of implicit Stokes 
(velocity-pressure) substeps per unit time interval.

Each Stokes substep requires the solution of 3 Helmholtz problems---one for each
velocity component---which are diagonally dominant and efficiently treated
by using Jacobi-preconditioned conjugate gradient iteration (Jacobi-PCG),
and a Poisson solve for the pressure, which bypasses the need to track
fast acoustic waves.  Because of long-range interactions that make
the problem communication intensive, the pressure Poisson solve is the 
dominant substep ($\approx$80\% of runtime) in the NS time-advancement.
To address this bottleneck, we use a variety of acceleration algorithms,
including $p$-multigrid (pMG) preconditioning, projection-based solvers, 
and projection-based initial guesses.

The tensor-product structure of spectral elements makes implementation of pMG
particularly simple.  Coarse-to-fine interpolations are cast as efficient tensor
contractions, $\uu_f = (\Jh \otimes \Jh \otimes \Jh) \uu_c$, where
$\Jh$ is the 1D polynmial interpolation operator from the coarse GLL points to
the fine GLL points.  Like differentiation, interpolation is on the reference
element, so only a single $\Jh$ matrix (of size $N+1$ or less) is needed for the
entire domain, for each pMG level.
   Smoothers for pMG include Chebyshev-accelerated point-Jacobi,
   additive Schwarz \cite{lottes05}, and Chebyshev-accelerated overlapping
   Schwarz.  
The Schwarz smoothers are implemented by solving local Poisson problems 
on domains extended into adjacent elements by one layer.  Thus, for an
$8\times 8 \times 8$ brick ($N=7$), one solves a $10\times 10 \times 10$
local problem, $\At^e \utu^e = \utb^e$,
using fast diagonalization (FDM) \cite{dfm02},
\begin{eqnarray} \label{eq:fdm}
\utu^e =
(S_z \otimes S_y \otimes S_x) \Lambda^{-1}
(S^T_z \otimes S^T_y \otimes S^T_x) \ub^e ,
\\[-3.0ex] \nonumber
\end{eqnarray} 
$e=1,\dots,E,$
where each 1D matrix of eigenvectors, $S_*$, is $10\times 10$,
and $\Lambda$ is a diagonal matrix with only $1,000$ nonzeros.
The leading complexity is $\approx 12 \times 10^4$ operations for
the application of $S_*$ and $S_*^T$ (implemented as {\tt dgemm}),
with $\approx 2,000$ loads per element
(for $\ub^e$ and $\Lambda$).  By constrast,
$A^e$, if formed, would have 1 million nonzeros
{\em for each element}, $\Omega^e$, which would make work and
storage prohibitive.   {\em The fast low-storage tensor decomposition
is critical to performance of the SEM}, as first noted in the seminal
paper of Orszag \cite{sao80}.
Note that the communication for the Schwarz solves is also very low.
We exchange {\em face data only} to get the domain extensions---meaning 6
exchanges per element instead of 26. This optimization yields a 10\%
speedup in runs at the strong-scale limit.   
(We have also implemented a {\em restricted additive Schwarz} (RAS) variant,
that does not require communication after the local solve, which cuts
communication of the smoother by a factor of two.)
Further preconditioner cost savings are realized by performing all steps of
pMG in 32-bit arithmetic.   On Summit, which has a limited number of
NICs per node, this approach is advantageous because it
reduces the off-node bandwidth demands by a factor of two. 

\medskip
\noindent
{\bf Projection Is Key}.
For incompressible flows, the pressure evolves smoothly in time, and 
one can leverage this temporal regularity by projecting known components
of the solution from prior timesteps.
For any $n \times L$ subspace of $\RR^n$ with $A$-orthonormal\footnote{Here 
$A$ is the discrete equivalent of $-\nabla^2$, which is symmetric positive
definite (SPD).}
basis $P_L=[ \upt_1  \dots \upt_L ]$, the {\em best-fit approximation} 
to the solution of $A \up^m =\ub^m$ is $\upb=P_LP_L^T \ub^m$, which 
can be computed with a single all-reduce of length $L$ ($\lesssim 30$).
The residual for the reduced problem $A \udp=\ub^m-A\upb$ has a significantly
smaller norm such that relatively few GMRES iterations are required to
compute $\udp$.  We augment the space $P_{L+1} = [ P_L \;\; \udp]$
after orthonormalizing $\udp$ against $P_L$ with one round of classical
Gram-Schmidt orthogonalization, which requires only a single all-reduce.  
(This approach is stable because $\udp$ is nearly $A$-orthogonal to $P_L$.)
In Sec. 7, we take $L_{\max}=30$ before 
restarting the approximation space, which yields a  1.7 speedup 
in NS solution time compared to $L$=0.

We remark that this projection algorithm is one of the {\em few instances
in distributed-memory computing} where one can readily leverage the 
additional memory that comes with increasing the number of ranks, $P$.
For low rank counts, one cannot afford $30$ vectors (each of size
$n/P$ per rank) and must therefore take $L < 30$, which results in
suboptimal performance, as observed in the strong-scaling results of
Sec.\ref{sec5}.
{\it With increasing $P$, this solution algorithm
improves because more memory is available for projection.}


Subsequent to projection, we use (nonsymmetric) pMG-preconditioning applied to
either flexible conjugate gradients (FlexCG) or GMRES.  FlexCG uses a short
recurrence and thus has lower memory and communication demands, whereas GMRES must
retain $K$ vectors for $K$ iterations.  With projection and a highly optimized
preconditioner, however, we need only $K \lesssim 5$, which keeps both memory and 
work requirements low while retaining the {\em optimal} projective properties
of GMRES.  For the full NS solution with $n$=51B, GMRES outperforms FlexCG by
about 3.3\% on $P=27648$ V100s.

\medskip
\noindent{\bf Autotuned Communication.}
On modern GPU platforms, one needs $n_{0.8}$$\approx$2.5 M points per GPU to
realize $\approx$ 80\% efficiency \cite{nekrs}.  For $N=7$, this implies 
$E_p$$\approx$7300 elements per MPI rank (GPU), which provides significant
opportunity for overlapping computation and communication because the majority 
of the elements are in the subdomain interior.  NekRS supports several 
communication strategies for the SEM gather-scatter ({\em gs}) operation, 
which requires an exchange-and-sum of shared interface values:
{\em (a)} pass-to-host, pack buffers, exchange; 
{\em (b)} pack-on-device, copy to host, exchange; and
{\em (c)} pack-on-device, exchange via GPUDirect.
They can be configured to overlap nonlocal {\em gs} communication with
processor-internal {\em gs} updates and other local kernels (if applicable). 

During {\em gs-setup}, the code runs a trial for each scenario and picks the
fastest opetion.   We note that processor adjacency is established 
in $\log P$ time through Nek5000's {\em gslib} utility, which has 
scaled to billions of points on millions of ranks.  The user (or code)
simply provides a global ID for each vertex, and {\em gslib} identifies
an efficient communication pattern to effect the requisite exchange.
(For $n=51$B, setup time on all of Summit is 2.8 s.)

The computational mesh can have a profound influence on solution quality 
and on the convergence rate of the iterative solvers.   We developed a 
code to generate high-quality meshes for random-packed spherical beds 
that was used in all pebble-bed cases shown.
   It is based on a tessellation of Voronoi cells with sliver removal
to generate isotropic mesh distributions.  The tessellated Voronoi
facets are projected onto the sphere surfaces to sweep out hexahedral
subdomains that are refined in the radial direction to provide 
boundary-layer resolution.  This approach yields about 1/6 the number
of elements as earlier tet-to-hex conversion approaches, which allows
us to elevate the local approximation order for the same resolution
while providing improved conditioning and less severe CFL constraints.

%% file: tex/measurement.tex
\section{Performance Measurements} 

\begin{table} 
 \footnotesize
\begin{center} \begin{tabular}{|l|c|c|c|c|}
  \hline
  \multicolumn{5}{|c|}{{\bf NekRS Timing Breakdown: n=51B, 2000 Steps}}\\
  \hline
& \multicolumn{2}{c}{{pre-tuning}}& \multicolumn{2}{|c|}{{post-tuning}} \\ \hline
Operation  & time (s) & \%  & time (s) & \%   \\ \hline
  computation        & 1.19+03  & 100 & 5.47+02 & 100 \\
  advection          & 5.82+01  &   5 & 4.49+01 &   8 \\
  viscous update     & 5.38+01  &   5 & 5.98+01 &  11 \\
  pressure solve     & 1.08+03  &  90 & 4.39+02 &  80 \\ \hspace{.1in}
          precond.   & 9.29+02  &  78 & 3.67+02 &  67 \\ \hspace{.3in}
      coarse grid    & 5.40+02  &  45 & 6.04+01 &  11 \\
  projection         & 6.78+00  &   1 & 1.21+01 &   2 \\
  dotp               & 4.92+01  &   4 & 1.92+01 &   4 \\ \hline
 \end{tabular}
\end{center}
\caption{\label{tab:break1} Default NekRS statistics output, provided
every 500 timesteps for each run, each user.  This table shows
results for the 352K pebble geometry of Fig. 1 on $P$=27648 V100s
on Summit.
}
\end{table}

Our approach to performance assessment is both top-down and bottom-up.
The top-down view comes from extensive experience on CPUs and
GPUs of monitoring timing breakdowns for large-scale Navier-Stokes
simulations with Nek5000 and NekRS.  The bottom-up perspective is
provided through extensive GPU benchmarking experience of the 
OCCA/libParanumal team (e.g.,\cite{occa,warburton2019}).

Every NekRS job tracks basic runtime statistics using a combination of
MPI\_Wtime and cudaDeviceSynchronize or CUDA events.  These are output every
500 time steps unless the user specifies otherwise.  Timing breakdowns roughly
follow the physical substeps of advection, pressure, and viscous-update, plus
tracking of known communication bottlenecks.
Table \ref{tab:break1} illustrates the standard output and
how it is used in guiding performance optimization.
We see a {\em pre-tuning} case that used default settings
(which might be most appropriate for smaller runs).
This case, which corresponds to 352K-pebble case ($E$=98M, $N$=8) 
on all of Summit indicates that 
45\% of the time is spent in the coarse-grid solve
and that the pressure solve constitutes 90\% of the 
overall solution time.   
Armed with this information and an understanding of multigrid, it is clear 
that a reasonable mitigation strategy is to increase the effectivenes of 
the smoothers at the higher levels of the pMG V-cycles.  As discussed in
Sec. 7, this indeed was a first step in optimization---we switched from
Chebyshev-Jacobi to Chebyshev-Schwarz with 2 pre- and post-smoothings,
and switched the level schedule from $N=8$, 5, 1 to $N=8$, 6, 4, 1,
where $N=1$ corresponds to the coarse grid, which is solved using Hypre
on the CPUs.   Additionally, we see in the pre-tuning column that 
projecting the pressure onto $L=8$ previous-timestep solutions 
accounted for $<$ 1 s of the compute time, meaning that we could readily
boost dimension of the approximation space, $R(P_L)$ to $L=30$.
With these and other optimizations, detailed in the next section, 
the solution time is reduced by a factor of two, as seen in the
{\em post-tuning} column.  In particular, we see that the coarse-grid
solve, which is a perennial worry when strong-scaling (its cost must
scale at least as $\log P$, rather than $1/P$ \cite{fischer15,tufofisc01}),
is reduced to a tolerable 11\%.

 \begin{table} [!t]
 \footnotesize
 \begin{center} \begin{tabular}{|l|c|c|c|c|c|c|c|}
   \hline
   \multicolumn{8}{|c|}{NVIDIA$\circledR$  Nsight\textsuperscript{TM} Compute Profiling }\\
   \multicolumn{8}{|c|}{{\bf NekRS Timing BreakDown: n=51B, 2000 Steps }}\\
   \multicolumn{8}{|c|}{{27,648 GPUs, $n/P=1.8M$, $E/P=3573$, $N=8$, $N_{q}=11$ }}\\
   \hline
   kernel           & time       &{\tt SD}     &{\tt SL}    &{\tt SM}    &{\tt PL}    &{\tt RL} & {\tt TF} \\
                    & [$\mu s$]  &[\%]         &[\%] &[\%] &      &[\%]   &     \\
   \hline
  $p$MG8            &      &     &     &     &    &     & FP32       \\
  Ax                &  144 &74   &59   &53   &G   &80   &2.34       \\
  gs                &  47  &64   &55   & -   &G   &70   &-         \\
  fdm               &  171 &54   &86   &72   &S   &98   &3.90       \\
  gs$_{\rm ext}$    &  68  &74   &36   & -   &G   &80   &-         \\
  fdm$_{\rm apply}$ &  83  &77   &26   & -   &G   &84   &-         \\
 \hline
  $p$MG6            &      &     &     &     &    &     & FP32        \\
  Ax                &  66  &75   &59   &50   &G   &82   &1.87       \\
  gs                &  28  &50   &56   &-    &L   &64   & -         \\ 
  fdm               &  81  &40   &86   &24   &S   &98   &3.12       \\
  gs$_{\rm ext}$    &  35  &64   &56   &-    &L   &70   & -           \\
  fdm$_{\rm apply}$ &  40  &72   &42   &-    &G   &78   &   -       \\
 \hline
  $p$MG4            &      &     &     &     &    &     & FP32        \\
  Ax                &  29  &65   &47   &34   &G   &71   &1.25       \\
  gs                &  16  &29   &50   &-    &L   &57   &  -          \\ 
  fdm               &  49  &45   &76   &60   &S   &86   &2.18       \\
   gs$_{\rm ext}$   &  21  &64   &56   &-    &L   &70   &  -          \\
  fdm$_{\rm apply}$ &  21  &67   &44   &-    &G   &73   & -         \\
 \hline
  CHAR              &      &     &     &     &    &     &FP64       \\
  adv               &1250  &60   &63   &36   &S   &72   &2.50       \\
  RK                & 250  &84   &-    &-    &G   &91   &  -         \\
  gs                &  88  &68   &36   &-    &G   &74   & -         \\
 \hline
 \end{tabular}
 \end{center}
   \caption{\label{prof2}
    Kernel analysis for 352K pebbles simulations on Summit at full-system scale.
    {\tt SD}:= SOL DRAM, 
    {\tt SL}:= SOL L1, 
    {\tt SM}:= SM Utilizaion, 
    {\tt PL}:= Performance Limiter, 
    {\tt RL}:= Roofline Performance, 
    {\tt TF}:= TFLOPS. [$\cdot$] is unit and `-' represents n/a.
     $p${\tt MG8}, 
     $p${\tt MG6}, 
     $p${\tt MG4}, 
     represent the kernels for $p$ multigrid with smoothing 
     for the degrees of polynomials 8,6,4, respectively.
    {\tt gs}:= All Other Near-Neighbor Updates (26 msgs per element),
    {\tt gs}$_{\rm ext}$:= Overlapping Schwarz Exchange (6 mgs per element),
    {\tt fdm}:= Fast Diagonalization Method, Eq. (3),
    and {\tt fdm}$_{\rm apply}$:= Update the solution with contributions from overlap.
    $n/P$: number of grid points per GPU,
    $E/P$: number of elements per GPU,
    $N$: polynomial order, and
    $N_q$: number of quadrature points for advection operator.
   }
 \end{table}

 \begin{table*}
 \footnotesize
 \begin{center} \begin{tabular}{|r|r|r|r|r|r|l|}
   \hline
   \multicolumn{7}{|c|}{Compute Profiling GPU Time Breakdown (43\% of Run Time)}\\
   \multicolumn{7}{|c|}{27,648 GPUs, $n/P=1.8M$, $E/P=3573$, $N=8$, $N_{q}=11$ }\\
   \hline
time [\%]  & total time    & instances  & average  & min & max & name \\
   \hline
 11.8&  37455851267&      63984 &585394         &530748 &688475 &\_occa\_nrsSubCycleStrongCubatureVolumeHex3D\_0\\
 11.2&  35299998642&    1003791 &35166.7        &8704   &76128  &\_occa\_gatherScatterMany\_floatAdd\_0          \\
 9.5 &  29998225559&    523524  &57300.6        &24223  &112223 &\_occa\_fusedFDM\_0                            \\
 8.2 &  25841854995&    523713  &49343.5        &16320  &163294 &\_occa\_ellipticPartialAxHex3D\_0              \\
 6.5 &  20589069440&    812325  &25345.9        &5696   &79296  &\_occa\_scaledAdd\_0                           \\
 6.3 &  20084031405&    159758  &125715.3       &8672   &338654 &\_occa\_gatherScatterMany\_doubleAdd\_0         \\
 4.1 &  12813250728&    1003791 &12764.9        &5823   &25504  &\_occa\_unpackBuf\_floatAdd\_0                  \\
 4.0 &  12785871394&    261762  &48845.4        &20992  &83487  &\_occa\_postFDM\_0                             \\
 3.8 &  12026221342&    1003791 &11980.8        &6016   &30688  &\_occa\_packBuf\_floatAdd\_0                    \\
 3.0 &  9498908211 &     31992  &296915.1       &234334 &486589 &\_occa\_nrsSubCycleERKUpdate\_0                \\
   \hline
\end{tabular}
 \end{center}
   \caption{\label{prof1} GPU time breakdown
   }
 \end{table*}

For GPU performance analysis, we use NVIDIA's profiling tools.
Table~\ref{prof2} summarizes the kernel-level metrics for the
critical kernels, which are identified with NVIDIA's Nsight Systems
and listed in Table \ref{prof1}.  The principal kernels are consistent
with those noted in Table \ref{tab:break1}, namely, the pMG smoother
and the characteristics-based advection update.
The kernel-level metrics (SOL DRAM, SOL L1/TEX Cache, SM utilization) 
of Table \ref{prof2} were obtained with NVIDIA's Nsight compute profiler.  
They indicate that the leading performance limiter (LPL) of most kernels
is the globlal memory bandwidth but in some cases the shared memory utilization
is also significant.  However two kernels are clearly shared memory bandwidth
bound.  {\it All kernels achieve a near roofline performance (RL) defined as
$>$70\% maximum realizable LPL utilization} (e.g. triad-STREAM produces 92\% of
GMEM) except the latency bound gather-scatter kernels in the coarse pMG levels.


%% file: tex/performance.tex
\section{Performance Results} 
\label{sec5}


 \begin{table}[!b]
 \footnotesize
 \begin{center} \begin{tabular}{|c|c|c|c|c|c|c|}
  \hline
  \multicolumn{7}{|c|}{{\bf NekRS Strong Scale:  Rod-Bundle,  200 Steps}}\\
  \hline
 Node & GPU   & $E$ & $n$ & $n/P$ & $t_{\rm step} [s] $ & Eff  \\
 \hline
 1810 &10860 &  175M & 60B &5.5M &1.85e-01 & 100  \\
 2536 &15216 &  175M & 60B &3.9M &1.51e-01 &  87  \\
 3620 &21720 &  175M & 60B &2.7M &1.12e-01 &  82  \\
 4180 &25080 &  175M & 60B &2.4M &1.12e-01 &  71  \\
 4608 &27648 &  175M & 60B &2.1M &1.03e-01 &  70  \\
  \hline
 \hline
  \multicolumn{7}{|c|}{{\bf NekRS Weak Scale: Rod-Bundle, 200 Steps}}\\
  \hline
 Node & GPU &  $E$ & $n$      &  $n/P$& $t_{\rm step}[s]$ & Eff \\
 \hline
 87   & 522   & 3M        & 1.1B  &  2.1M  & 8.57e-02  & 100   \\
 320  & 1920  & 12M       & 4.1B  &  2.1M  & 8.67e-02  & 99    \\
 800  & 4800  & 30M       & 10B   &  2.1M  & 9.11e-02  & 94    \\
 1600 & 9600  & 60M       & 20B   &  2.1M  & 9.33e-02  & 92    \\
 3200 & 19200 & 121M      & 41B   &  2.1M  & 9.71e-02  & 88    \\
 4608 & 27648 & 175M      & 60B   &  2.1M  & 1.03e-01  & 83    \\
 \hline
 \end{tabular}
\end{center}
 \caption{\label{rod-strong-weak}
    NekRS strong and weak scaling for rod bundle simulations.
    $n/P$: number of grid points per gpu,
    $E/P$: number of elements per gpu,
    $t_{\rm step}$: average wall time for 101--200 steps,
    and Eff: efficiency.
    BDF3+EXT3 is used for timestepping with $\Delta t=$ 3e-4, 
    corresponding to CFL=0.54.
   }
\end{table}

\medskip
\noindent
{\bf 17$\times$17 Rod Bundle.}
We begin with strong- and weak-scaling studies of a 17$\times$17 pin bundle
of the type illustrated in Fig. \ref{fig:350k} (far right).  The mesh 
comprises 27700 elements in the $x$-$y$ plane and is extruded in 
the axial ($z$) direction.  Any number of layers can be chosen in $z$
in order to have a consistent weak-scale study.
In this study, we do not use characteristics-based timestepping,
but instead use a more conventional semi-implicit scheme that
requires CFL $\lesssim$ 0.5 because of the explicit treatment
of the nonlinear advection term.    The initial condition is
a weakly chaotic vortical flow superimposed on a mean axial
flow and timings are measured over steps 100--200.
Table \ref{rod-strong-weak}(top) presents strong scale results on
Summit for a case with $E=175$M and $N=7$  ($n$=60B) for $n/P$ ranging
from 5.5M down to 2.1M.  We see that 80\% efficiency is realized
at $n/P \approx 2.5$M.   The weak-scale study, taken at a 
challenging value of $n/P=2.1$M, shows that our solvers sustain
up to 83\% (weak-scale) parallel efficiency out to the full 
machine.

We measured the average wall time per step in seconds, $t_{step}$, using
101-200 steps for simulations with $Re_D=5000$.  The approximation order is
$N=7$, and dealiasing is used with $N_q=9$.  We use projection in time,
CHEBY+ASM, and flexible PCG for the pressure solves with tolerance 1.e-04.  The
velocity solves use Jacobi-PCG with tolerance 1.e-06. BDF3+EXT3 is used for
timestepping with $\dt$= 3.0e-04, corresponding to CFL=0.54.
The pressure iteration counts, $p_i \sim$ 2, are lower for these cases
than for the pebble cases, which have $p_i\sim$ 8 for the same timestepper
and preconditioner.  The geometric complexity of the rod bundles is relatively
mild compared to the pebble beds. Moreover, the synthetic initial condition
does not quickly transition to full turbulence.  We expect more pressure
iterations in the rod case (e.g., $p_i \sim$ 4--8) once turbulent 
flow is established.

\medskip
\noindent
{\bf Pebble Bed--Full Core.}  The main target of our study is the full core for
the pebble bed reactor (Fig. 1, center), which has 352,625 spherical pebbles
and a fluid mesh comprising $E=98782067$ elements of order $N=8$ ($n\approx
51$B).  In this case, we consider the characteristics-based timestepping
with $\dt=4$.e-4 or 8.e-4, corresponding to respective Courant numbers 
of $CFL \approx 2$ and 4.  Table \ref{tab:tmp} lists the battery of 
tests considered for this problem, starting with the single-sweep
Chebyshev-Additive Schwarz (1-Cheb-ASM) pMG smoother, which is the default
choice for smaller (easier) problems.  As noted in the preceding section, this
choice and the two-smoothings Chebyshev-Jacobi (2-Cheb-Jac) option yield very
high coarse-grid solve costs because of the relative frequency in which the
full V-cycle must be executed.   
Analysis of the standard NekRS output suggested that more smoothings
at the finer levels would alleviate the communication burden incurred
by the coarse-grid solves.  We remark that, on smaller systems, where
the coarse-grid solves are less onerous, one might choose a different
optimization strategy.  

\begin{figure}
  \centering
  \includegraphics[width=0.45\textwidth]{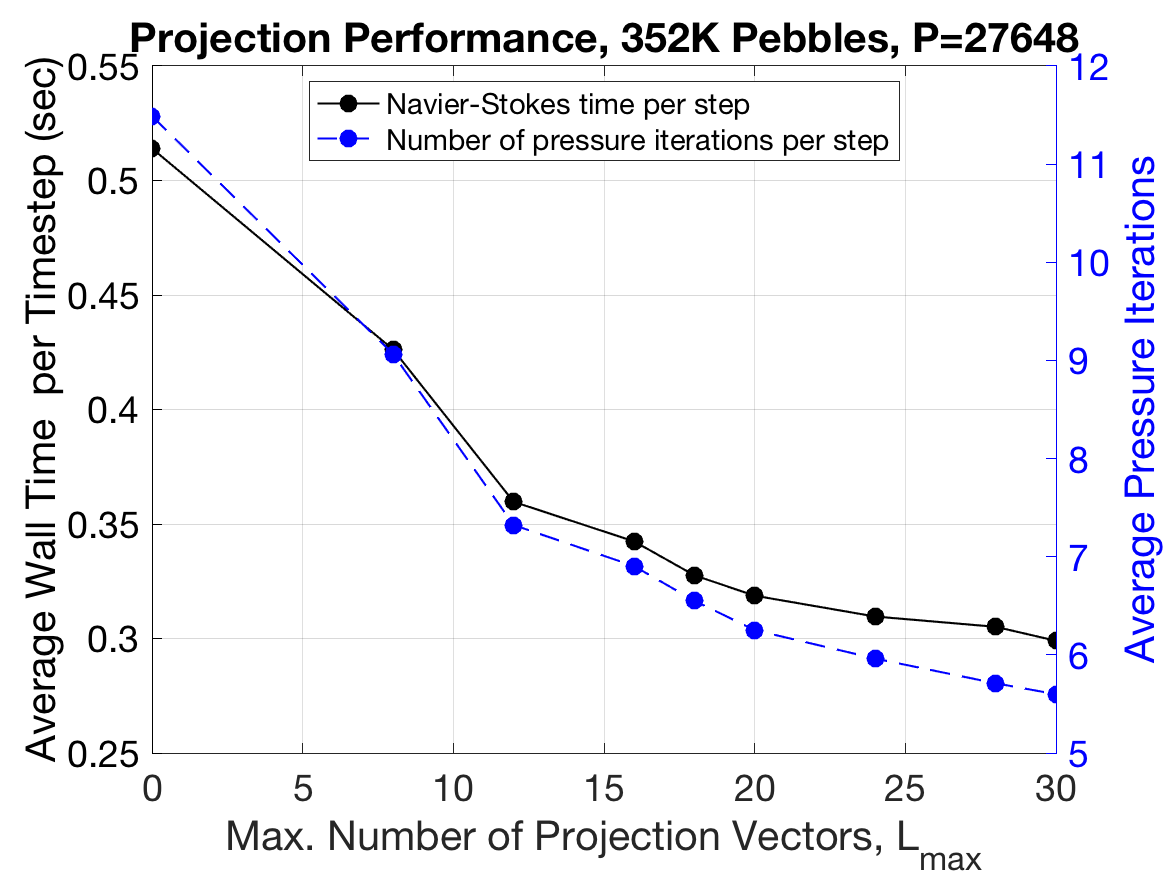}
  \caption{\label{fig:bk5} 
    Average time per step and pressure iteration count 
    as a function of projection-space dimension, $L_{\max}$.}
 \end{figure}

The first step in optimization was thus to increase the number of 
smoothings (2-Cheb-ASM) and to increase the number of pMG levels to 
four, with $N$=8, 6, 4, 1 (where 1 is the coarse grid).  These steps
yielded a 1.6$\times$ speed-up over the starting point.
  Subsequently, we boosted, $L$, the number of prior solutions to
use as an approximation space for the projection scheme from 8 to 30,
which yielded an additional factor of 1.7, as indicated in Fig. \ref{fig:bk5}.

  Given the success of projection and additional smoothing, which lowered
the FlexCG iteration counts to $<$6, it seemed clear that GMRES would be
viable.  A downside of GMRES is that the memory footprint scales as $K$, 
the maximum number of iterations and the work (and potentially, communication)
scales as $K^2$.  With $K$ bounded by 6, these complexities are not onerous 
and one need not worry about losing the projective property of GMRES by having 
to use a restarted variant.  Moreover, with so few vectors, the potential of
losing orthogonality of the Arnoldi vectors is diminished, which means that
classical Gram-Schmidt can be used and one thus has only a single all-reduce 
of a vector of length $< K$ in the orthogonalization step.

The next optimizations were focused on the advection term.
First, we reduced the number of quadrature points from $N_q=13$ 
to 11 (in each direction).  Elevated quadrature is necessary for
{\em stability}, but not for accuracy.  While one can prove stability
for $N_q \ge 3N/2$ \cite{johan13}, it is not mandatory and, when
using the characteristics method, which visits the advection operator
at least four times per timestep, it can pay to reduce $N_q$ as long
as the flow remains stable.   Second, we {\em increased} $\dt$ by a
factor of 2, which requires 2 subcycles to advance the hyperbolic
advection operator (i.e., doubling its cost), but does not double
the number of velocity and pressure iterations.
Case (f) in Table \ref{tab:tmp} shows that the effective cost 
(based on the original $\dt$) is $t_{\rm step}=0.188$ s.  In case (g)
we arrive almost at $t_{\rm step}=0.18$ s by turning off all I/O
to stdout for all timesteps modulo 1000.  
 The net gain is {\bf a factor of 3.8} over the starting (default)
point.

\medskip
\noindent
{\bf Pebble Bed Strong-Scale.}
Table \ref{peb35k-strong} shows three strong scale studies for the pebble bed
at different levels of optimization, with the final one corresponding
to Case (f) of Table \ref{tab:tmp}.  The limited memory on the GPUs means
that we can only scale from $P=9216$ to 27648 for these cases and in fact
cannot support $L=30$ at $P=9216$, which is why that value is absent from
the table.  

As noted earlier, a fair comparison for the last set of entries
would be to run the $P=9216$ case with a smaller value of $L$---it would 
perform worse, which would give a scaling advantage to the $L=30$ case.
This advantage is legimate, because $L=30$ is an improved algorithm
over (say) $L=8$, which leverages the increase memory resources that
come with increasing $P$.

\medskip
{\bf Rapid Turn-Around.}
Remarkably, the performance results presented here imply that it is now
possible to simulate a single flow-through time for a full core in less than 
six hours when using all of Summit for these simulations.   The analysis
proceeds as follows.  In the nondimensional units of this problem, the domain
height is $L_z=130$ units, while the mean flow speed in the pebble region is 
$U \approx 1/\phi$, where $\phi \approx 0.36$ is the void fraction in the packed
bed.  The nondimensional flow-through time is thus $L_z/U \approx 130 \times
0.36 = 46.6$ (modulo a reduction of the flow speed in the upper and lower
plena).  The timestep size from the final two rows of Table 5 is $8 \times
10^{-4}$, corresponding to 58250 steps for a single flow-through time, which,
at 0.361 seconds per step, corresponds to 5.84 hours of wall-clock time per
flow-through time.

\begin{table*} 
\footnotesize
\begin{center} \begin{tabular}{|l|c|c|c|c|c|c|c|l|l|}
  \hline
\multicolumn{9}{|c|}{{\bf Major Algorithmic Variations, 352K Pebbles, $P$=27648}}\\ \hline
{\rm Case} & Solver & Smoother           & $L$&$N_q$&$\dt$ &$v_i$ & $p_i$ & $t_{\rm step}$  \\ \hline
(a)        & FlexCG &1-Cheb-ASM:851      &  8 &  13 & 4e-4 &  3.6 & 22.8  & .68  \\
(b)        & FlexCG &2-Cheb-Jac:851      &  8 &  13 & 4e-4 &  3.6 & 17.5  & .557 \\
(c)        &  "     &2-Cheb-ASM:851      &  8 &  13 & 4e-4 &  3.6 & 12.8  & .468 \\
(d.8)      &  "     &2-Cheb-ASM:8641     &  8 &  13 & 4e-4 &  3.6 &  9.1  & .426 \\
(d.$L$)    &  "     &2-Cheb-ASM:8641     &0--30& 13 & 4e-4 &  3.6 &  5.6  & .299 \\ 
(e)        & GMRES  &    "               &  30&  13 & 4e-4 &  3.5 &  4.6  & .240 \\ \hline
(f)        &  "     &    "               &  30&  11 & 8e-4 &  5.7 &   7.2 & .376 \\
(g)        &  "     &    "               &  30&  11 & 8e-4 &  5.7 &   7.2 & .361 (no I/O)   \\
 \hline  
 \end{tabular}
\end{center}
  \caption{\label{tab:tmp}  Progression of algorithmic trials. 
See Fig. \ref{fig:bk5} for Cases (d.$L$), $L$=0:30.
   }
\end{table*}

\begin{table}
\footnotesize
\begin{center} 
\begin{tabular}{|c|c|c|c|c|c|c|}
  \hline
  \multicolumn{7}{|c|}{{\bf NekRS Strong Scale: 352K pebbles, $E$=98M, $n$=50B}}\\
  \hline
  \multicolumn{7}{|c|}{{\bf$N=8$, $N_q=13$, $\Delta t=$ 4.e-4, $L=8$, 1-Cheb-Jac:851}}\\
  \hline
  Node & GPU  & $n/P$ &  $v_i$ & $p_i$ & $t_{\rm step}$ & Eff   \\
  \hline
  1536   &  9216  &   5.4M    &  3.6 &   17.3&   .97  &   1.00      \\
  2304   &  13824 &   3.6M    &  3.6 &   18.0&   .84  &   76.9      \\
  3072   &  18432 &   2.7M    &  3.6 &   16.6&   .75  &   64.6      \\
  3840   &  23040 &   2.1M    &  3.6 &   19.6&   .67  &   57.9      \\
  4608   &  27648 &   1.8M    &  3.6 &   17.5&   .55  &   58.7      \\
 \hline
 \hline
  \multicolumn{7}{|c|}{{\bf $N=8$, $N_q=13$, $\Delta t=$ 4.e-4, $L=8$, 1-Cheb-ASM:851}}\\
  \hline
  Node & GPU  & $n/P$  & $v_i$ & $p_i$ & $t_{\rm step}$ & Eff  \\ 
  \hline
  1536   &  9216  &   5.4M    &  3.6 &   11.6 &   .81&    100       \\
  2304   &  13824 &   3.6M    &  3.6 &   12.3 &   .65&    83.0      \\
  3072   &  18432 &   2.7M    &  3.6 &   12.3 &   .71&    57.0      \\
  3840   &  23040 &   2.1M    &  3.6 &   13.5 &   .54&    60.0      \\
  4608   &  27648 &   1.8M    &  3.6 &   12.8 &   .46&    58.6      \\
 \hline
 \hline
  \multicolumn{7}{|c|}{{\bf $N=8$, $N_q=11$, $\Delta t=$ 8.e-4, $L=30$, 2-Cheb-ASM:8641}}\\
  \hline
  Node & GPU  & $n/P$ &  $v_i$ & $p_i$ & $t_{\rm step}$ & Eff \\
  \hline
  1536   &  9216  &   5.4M   &  -   &   -  &    -   &      -       \\
  2304   &  13824 &   3.6M   &   5.7&   7.2&    .55 &    100       \\
  3072   &  18432 &   2.7M   &   5.7&   7.2&    .56 &    73.6      \\
  3840   &  23040 &   2.1M   &   5.7&   7.2&    .39 &    84.6      \\
  4608   &  27648 &   1.8M   &   5.7&   7.2&    .36 &    76.3      \\
 \hline
 \end{tabular}
\end{center}
 \caption{\label{peb35k-strong}
 NekRS Strong Scale using BDF2 with characteristic.}
\end{table}

%% file: tex/conclusion.tex
\section{Conclusions} 

The simulations of full nuclear reactor cores described here are ushering in a
new era for the thermal-fluids and coupled analysis of nuclear systems. The
possibility of simulating such systems in all their size and complexity was
unthinkable until recently. In fact, the simulations are already being used to
benchmark and improve predictions obtained with traditional methods such as
porous media models. This is important because  models currently in use were not
developed to predict well the change in resistance that occurs in the cross
section due to the restructuring of the pebbles in the near wall region. This
is a well known gap hindering the deployment of this class of reactors.  Beyond
pebble beds, the fact that such geometry can be addressed with such low
time-to-solution will enable a broad range of optimizations and reductions in
uncertainty in modeling that were until now not achievable in nuclear
engineering.  The impact will extend to all advanced nuclear reactor design
with the ultimate result of improving their economic performance. This will in
turn serve broadly the goal of reaching a carbon-free economy within the next
few decades.

The study presented here demonstrates the continued importance of numerical
{\em algorithms} in realizing HPC performance, with up to a four-fold reduction
in solution times realized by careful choices among a viable set of options.  
This optimization was realized in relatively short time (a matter of days) 
by having a suite of solution algorithms and implementations available
in NekRS---no single strategy is always a winner.  For users, who often have a
singular interest, being able to deliver best-in-class performance can make all
the difference in productivity.   In Nek5000 and NekRS, we support
automated tuning of communication strategies that adapt to the network and
underlying topology of the particular graph that is invoked at runtime.  This
approach has proven to make up to a factor of 4 difference, for example, in AMG
implementations of the coarse-grid solver.

%% file: tex/acknowledge.tex
\section*{Acknowledgments}

This material is based upon work supported by the U.S. Department of Energy, 
Office of Science, under contract DE-AC02-06CH11357 
   and by the Exascale Computing Project (17-SC-20-SC), a
collaborative effort of two U.S. Department of Energy organizations (Office of
Science and the National Nuclear Security Administration) responsible for the
planning and preparation of a capable exascale ecosystem, including software,
applications, hardware, advanced system engineering and early testbed platforms,
in support of the nation's exascale computing imperative.

   The research used resources at the Oak Ridge Leadership Computing Facility
at Oak Ridge National Laboratory, which is supported by the Office of Science
of the U.S. Department of Energy under Contract DE-AC05-00OR22725 and at the
Argonne Leadership Computing Facility, under Contract DE-AC02-06CH11357. 
